\documentclass[10pt,conference]{IEEEtran}

\usepackage{amsmath}
\usepackage{balance}
\usepackage[hyphens]{url}
\usepackage{booktabs}
\usepackage{hyperref}
\usepackage{graphicx}
\usepackage{soul}

\makeatletter
\newcommand{\linebreakand}{
  \end{@IEEEauthorhalign}
  \hfill\mbox{}\par
  \mbox{}\hfill\begin{@IEEEauthorhalign}
}
\makeatother

\title{Challenge on Optimization of Context Collection for Code Completion}

\author{
\IEEEauthorblockN{Dmitry Ustalov}
\IEEEauthorblockA{
    \textit{JetBrains}\\
    Belgrade, Serbia\\
    dmitry.ustalov@jetbrains.com
}
\and
\IEEEauthorblockN{Egor Bogomolov}
\IEEEauthorblockA{
    \textit{JetBrains Research}\\
    Amsterdam, The Netherlands\\
    egor.bogomolov@jetbrains.com
}
\linebreakand
\IEEEauthorblockN{Alexander Bezzubov}
\IEEEauthorblockA{
    \textit{JetBrains Research}\\
    Amsterdam, The Netherlands\\
    alexander.bezzubov@jetbrains.com
}
\and
\IEEEauthorblockN{Yaroslav Golubev}
\IEEEauthorblockA{
    \textit{JetBrains Research}\\
    Belgrade, Serbia\\
    yaroslav.golubev@jetbrains.com
}
\linebreakand
\IEEEauthorblockN{Evgeniy Glukhov}
\IEEEauthorblockA{
    \textit{JetBrains Research}\\
    Amsterdam, The Netherlands\\
    evgeniy.glukhov@jetbrains.com
}
\and
\IEEEauthorblockN{Georgii Levtsov}
\IEEEauthorblockA{
    \textit{Neapolis University Pafos, JetBrains}\\
    Pafos, Cyprus\\
    g.levtsov.1@nup.ac.cy
}
\and
\IEEEauthorblockN{Vladimir Kovalenko}
\IEEEauthorblockA{
    \textit{JetBrains Research}\\
    Amsterdam, The Netherlands\\
    vladimir.kovalenko@jetbrains.com
}
}

\begin{document}

\maketitle

\begin{abstract}
The rapid advancement of workflows and methods for software engineering using AI emphasizes the need for a systematic evaluation and analysis of their ability to leverage information from entire projects, particularly in large code bases. In this challenge on optimization of context collection for code completion, organized by JetBrains in collaboration with Mistral~AI as part of the ASE~2025 conference, participants developed efficient mechanisms for collecting context from source code repositories to improve fill-in-the-middle code completions for Python and Kotlin. We constructed a large dataset of real-world code in these two programming languages using permissively licensed open-source projects. The submissions were evaluated based on their ability to maximize completion quality for multiple state-of-the-art neural models using the chrF metric. During the public phase of the competition, nineteen teams submitted solutions to the Python track and eight teams submitted solutions to the Kotlin track. In the private phase, six teams competed, of which five submitted papers to the workshop.
\end{abstract}

\begin{figure}[t]
\includegraphics[width=\columnwidth]{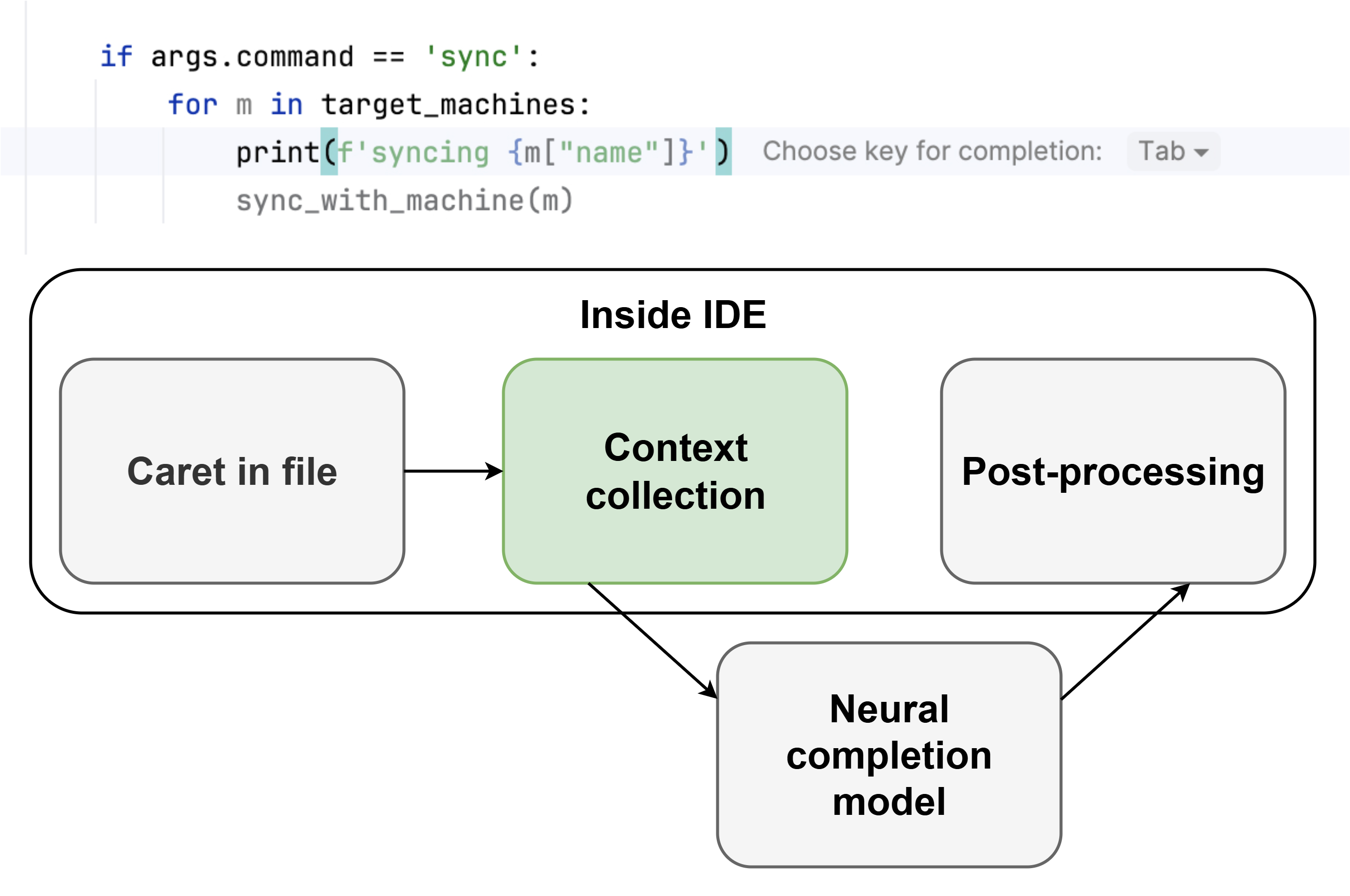}
\caption{\label{fig:overview}When a completion is requested, the IDE gathers the context at the caret position and generates the corresponding prompt for the neural completion model. The model's output is subsequently post-processed and presented as the suggested completion. In our competition, we aim to identify the most effective method for context collection (green block) while assuming that all other components remain unchanged.}
\vspace{-0.3cm}
\end{figure}

\section{Introduction}

Code completion is the task of predicting and inserting source code statements based on the current context~\cite{Izadi:22}. Most modern integrated development environments (IDEs) provide code completion functionality using neural networks, either through built-in models or external providers. Contemporary approaches to code completion adopt the \emph{fill-in-the-middle} formulation~\cite{Bavarian:22}, which involves infilling the code between a given prefix and suffix. In this setting, the prefix and suffix naturally correspond to the beginning and end of the code block currently being edited.

Recent academic publications~\cite{Zhang:23,Roziere:23,Wu:24} and our internal research~\cite{Bogomolov:24,Sapronov:25} indicate that the quality of neural code completions depends not only on the capabilities of the underlying model but also on \emph{the context}. By context, we mean the way in which file contents, symbol definitions, class hierarchies, and data types are collected and organized across the entire code base to supplement the prefix and suffix (Figure~\ref{fig:overview}). 

\begin{figure}[t]
\includegraphics[width=\columnwidth]{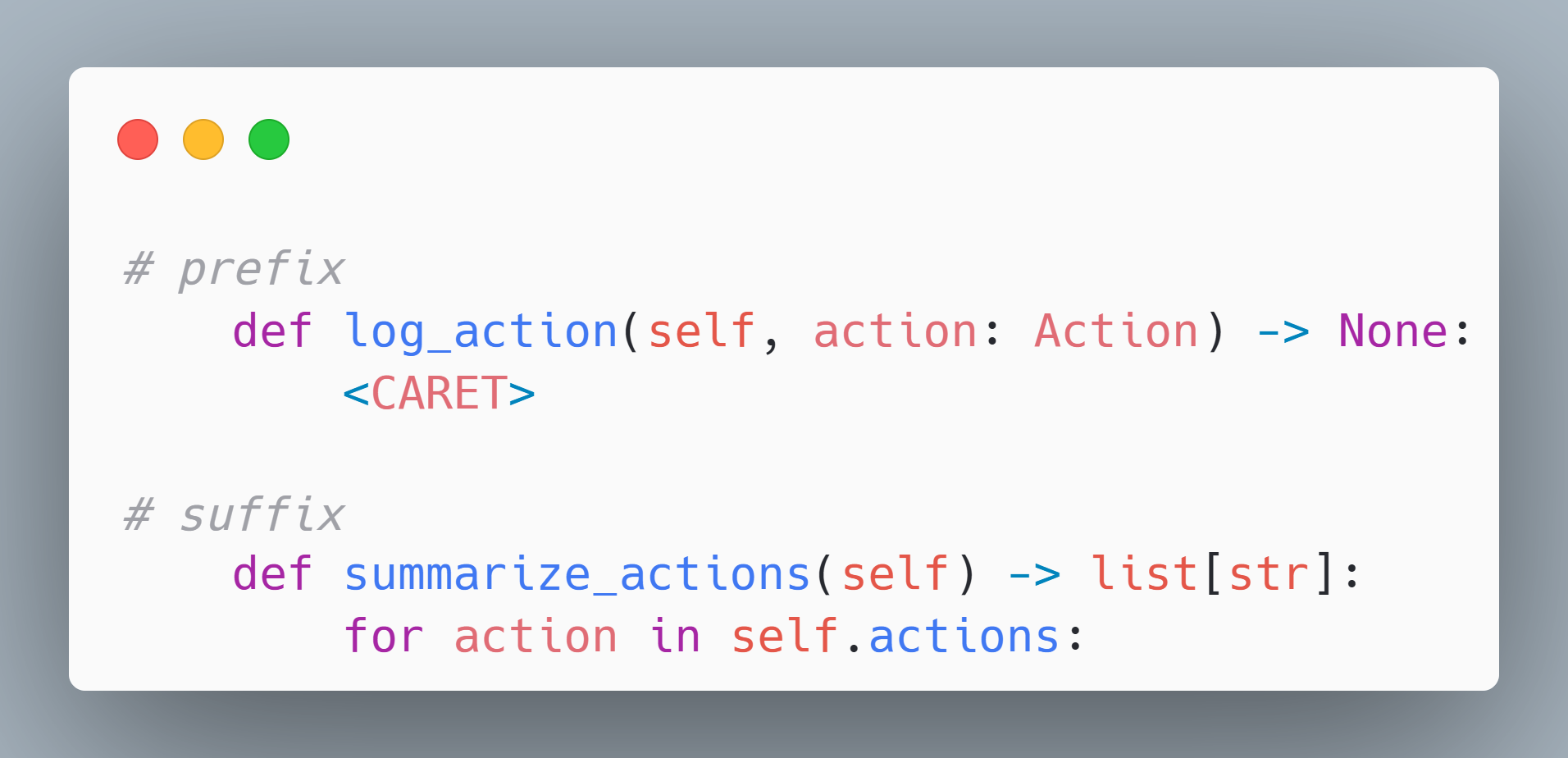}
\caption{The simplest context collection: no context clues, only some of the lines before and after the CARET (editing cursor) position in the given file.\label{fig:example:empty}}
\end{figure}

\begin{figure}[t]
\includegraphics[width=\columnwidth]{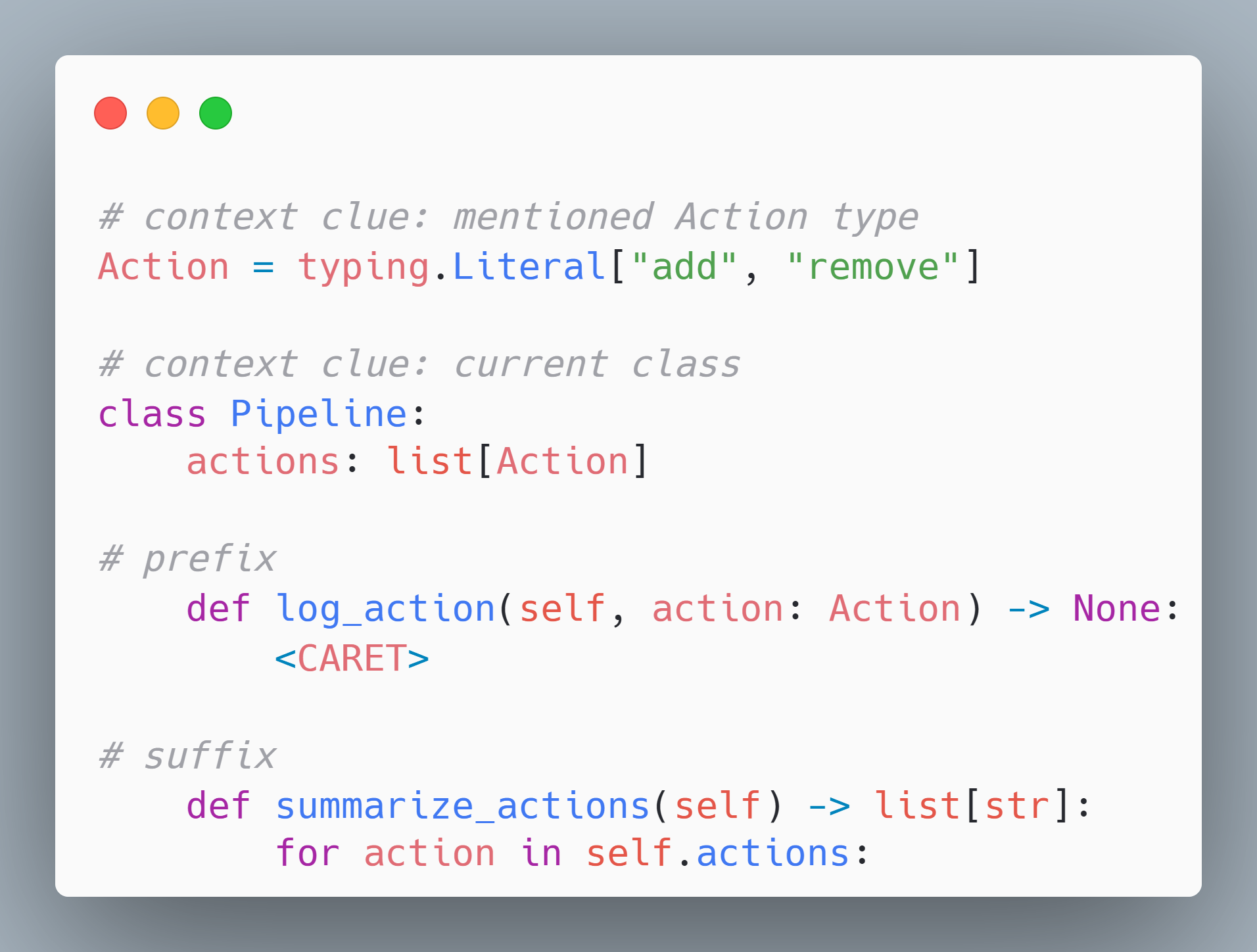}
\caption{The same completion point as in Figure~\ref{fig:example:empty}, but with context clues obtained from resolving the symbols mentioned in the code: \texttt{Action} type and the current class, \texttt{Pipeline}.\label{fig:example:context}}
\vspace{-0.3cm}
\end{figure}

Context quality is so critical that a smaller model with better \textit{context clues} can outperform a larger, more capable model~\cite{Wang:25}. Figures~\ref{fig:example:empty} and~\ref{fig:example:context} illustrate two inputs to a code completion model: one using only the lines surrounding the user's caret (Figure~\ref{fig:example:empty}) and another enriched with useful context clues (Figure~\ref{fig:example:context}). The latter example is expected to yield better completions, as the model itself has little inherent knowledge of the current project. Providing context clues allows the model to consider actual definitions and symbols during completion. Recent literature demonstrates non-trivial approaches to this problem, including applications of reinforcement learning~\cite{Wang:25} and information retrieval~\cite{Phan:24}, highlighting the complexity of the present task.

To foster research on this interdisciplinary problem at the intersection of natural language processing and software engineering, we organized a community competition focused on designing context collection strategies for AI-based code completion tools. The goal of the competition was to develop strategies for gathering context for code completion with large language models (LLMs), aiming to maximize the quality of the resulting fill-in-the-middle completions without exerting direct control over these completions. We created a new code completion dataset for two programming languages, Python and Kotlin, conducted multiple phases of the competition, and attracted more than fifteen teams from around the world to advance the state-of-the-art in this area.

The remainder of the paper is organized as follows. Section~\ref{sec:setup} defines the problem addressed in the challenge and describes the evaluation methodology we employed, along with the competition dataset we created. Section~\ref{sec:timeline} presents the competition timeline and the hosting procedure. Section~\ref{sec:solutions} provides an overview of the competition results and examines the solutions submitted during the private phase of the competition. Section~\ref{sec:conclusion} offers concluding remarks. 

\section{\label{sec:setup}Competition Setup}

The goal of our competition was to devise the most effective strategy for collecting context from the entire code base for the fill-in-the-middle code completion task. Given the prefix, suffix, caret position, and all files in the software project, which we collectively refer to as a \emph{completion point} (Figure~\ref{fig:example:empty}), participants needed to implement a context collector that outputs a string called \emph{the context}. This context should enable the best fill-in-the-middle completions (Figure~\ref{fig:example:context}) across three capable LLMs designed for coding tasks: Codestral by Mistral~AI~\cite{Codestral}, a popular open-source model Qwen2.5-Coder~\cite{Hui:24}, and our own model, Mellum~\cite{Mellum}. We evaluated the quality of the completions by comparing the contextualized outputs of the models with the ground truth; the overall quality was computed as the average across the three models.

The competition consisted of two tracks that shared the same problem definition but differed in their target programming languages and corresponding datasets:
\begin{itemize}
  \item the first track focused on \emph{Python}, which is a popular target for many AI-based programming assistance techniques due to its large and diverse user base;
  \item the second track focused on \emph{Kotlin},\footnote{Kotlin: \url{https://jetbrains.com/kotlin}} a language historically well supported in JetBrains products, but one that has attracted less attention in the research community.
\end{itemize}

Participants were invited to submit to both tracks. We were particularly interested in universal solutions capable of handling both a dynamically-typed language, Python, and a statically-typed language, Kotlin.

\subsection{Phases and Workflow}

Since the task in this competition is to implement only the context collector (Figure~\ref{fig:overview}), we followed the methodology used in the Toloka WSDM~Cup 2023 competition~\cite{TolokaWSDMCup2023} to ensure fairness, reproducibility of the results, and prevention of potential data leaks. We conducted our competition in three phases, held simultaneously for each of the two tracks:

\begin{itemize}
  \item \emph{Practice phase}, in which participants received the ground truth completions for a small subset of the competition dataset. They were also provided with a baseline notebook for local experiments to better understand the competition format.
  \item \emph{Public phase}, in which participants continued to receive ground truth completions and competed for the leaderboard. They had to surpass the baselines to qualify for the private phase.
  \item \emph{Private phase}, in which participants submitted the code of their solutions. We reviewed and ran the top solutions from the leaderboard on a separate held-out dataset with no Internet connection to determine the winners.
\end{itemize}

We followed the workflow below:

\begin{enumerate}
  \item the participant submitted the collected context for each completion point, not the actual neural completion;
  \item our competition platform accepted the submission and converted each context into a model-specific prompt for all three models by inserting the model-specific special tokens and arranging context, prefix, and suffix in the format required by the model;
  \item the platform requested completions, received the results, returned the evaluation scores for each completion, and calculated the average;
  \item the scores were displayed on the leaderboard on the platform during the public phase;
  \item the final rankings were determined during the private phase.
\end{enumerate}

During the public stage of the competition, participants were required to submit the data files without their code. They were allowed to use any content from the provided dataset. We provided a mechanism in the competition forums for participants to request permission to use external tools, such as Web search, in their solutions. After the public phase, we invited the authors of all solutions that outperformed the baselines to submit a working container image implementing their approach. In the private phase, we ran these containers on our machines using the held-out dataset, following the same protocol as in the public stage. The solutions that maximized the average $\mathrm{chrF}$ score on the private subset of the competition dataset were considered the winners of the challenge.

\vspace{0.2cm}

\subsection{Metric}

A study by Evtikhiev et al.~\cite{Evtikhiev:23} showed that a widely used evaluation criterion in machine translation, $\mathrm{chrF}$ (character F-score) \cite{Popovic:15}, is currently one of the most reliable indicators of code completion quality due to its interpretability and flexibility. The $\mathrm{chrF}$ score is defined as the harmonic mean of character-wise precision and recall, analogous to the F-score from information retrieval:

\begin{equation*}
  \mathrm{chrF} = 2 \frac{\mathrm{chrP} \cdot \mathrm{chrR}}{\mathrm{chrP} + \mathrm{chrR}}\text{,}
\end{equation*}
where $\mathrm{chrP}$ denotes the percentage of character $n$-grams in the suggestion that appear in the ground-truth completion, and $\mathrm{chrR}$ denotes the percentage of character $n$-grams in the ground-truth completion that also appear in the suggestion. We used the same implementation of $\mathrm{chrF}$ as the evaluation criterion throughout our competition, covering the practice, public, and private phases.

We selected several relatively strong baselines to establish a reasonable level of expectations, as it was necessary to run the code on-premise during the private phase of the competition. Based on our experience, most IDEs on the market use the \textit{recent files} strategy, which relies on the files currently opened in the editor. This strategy is not applicable in the offline setting of our competition. Since our dataset is derived from commits in large repositories of permissive code, we provide a proxy for this signal by listing other files changed in the same commit. This approach follows established practice in this research area~\cite{Eliseeva:23}. Additionally, we include a baseline strategy that selects the closest file according to the BM25~\cite{Robertson:09} ranking function, as well as a random context file strategy.

\subsection{Dataset}

We built our context collection competition on top of an existing benchmark, Long Code Arena (LCA)~\cite{Bogomolov:24}, which includes the single-line repository-level code completion task. This dataset simulates the way developers write code by using git commit histories to separate the file being modified from the repository snapshot used for context collection. It allows avoiding the possible temporal data leakages between the context and target completion, as the context comes from the repository snapshot present before the ground truth completion was written.

Following the approaches of LCA and Toloka~VQA~\cite{Bogomolov:24,TolokaWSDMCup2023}, we created a completely new dataset from permissively-licensed open-source repositories on GitHub, incorporating several important differences from LCA. Specifically, it provides multi-line completion using a fill-in-the-middle approach rather than prefix-based completion, it includes subsets for Python and Kotlin programming languages, and it separates the data into non-overlapping practice, public, and private subsets. We generated the multi-line splits using in-house code analysis tools employed in our IDEs, though we expect that the results of this processing are broadly applicable beyond our products. All code in our dataset is permissively licensed.

We collected only large code repositories, such as IPython\footnote{IPython: \url{https://github.com/ipython/ipython}} and dukat,\footnote{dukat: \url{https://github.com/Kotlin/dukat}} to allow the use of sophisticated context collectors on realistic code bases. We enumerated the complete history of commits and retained only those that included a significant number of multi-line insertions, identified using simple content-based heuristics. Each repository could contain one or more completion points, but no repository appeared in more than one subset of the data (practice, public, private).

Overall, the evaluation dataset consisted of 102 repositories, 1,176 revisions, and 1,764 completion points. For the Python track, there were 47 points in the practice phase, 247 points in the public phase, and 394 points in the private phase, totaling 688 points from 52 repositories. For the Kotlin track, there were 30 points in the practice phase, 400 points in the public phase, and 646 points in the private phase, totaling 1,076 points from 50 repositories. The dataset was separated once before the start of the competition, and it remained unchanged during the competition. The public dataset was available at the start of the competition, while the private dataset was released after the competition results were announced. The complete competition dataset, including ground truth data, repositories, and private phase submissions, can be downloaded from Zenodo~\cite{Dataset} under the CC~BY 4.0 license.\footnote{CC~BY 4.0: \url{https://creativecommons.org/licenses/by/4.0/}}

\section{\label{sec:timeline}Process and Timeline}

We hosted both the practice and public parts of the competition on the EvalAI platform~\cite{Yadav:19}.\footnote{EvalAI: \url{https://eval.ai/web/challenges/challenge-page/2516/overview}} Since our competition required collecting contexts for a given set of completion points and our evaluation protocol involved calling the neural completion models, we implemented these model calls on the platform side using API keys at our expense, with Codestral expenses offset by Mistral~AI. We also released a convenient starter kit\footnote{Starter Kit: \url{https://github.com/JetBrains-Research/ase2025-starter-kit}} to enable rapid onboarding and simplify prototyping, providing three baseline solutions: empty context, random recent file, and the most similar file according to BM25. Due to the use of multiple code completion models with different context lengths, we asked the participants to use the \texttt{<|file\_sep|>} separator in their outputs so we could take care of constructing the final prompt in a correct format for each model.

Table~\ref{tab:timeline} presents the competition timeline. In April, we deployed both tracks for internal testing and invited our colleagues at JetBrains to attempt the public part of the competition to identify potential mistakes, inconveniences, and documentation issues. JetBrains employees and their affiliates were not eligible for prizes under any circumstances, in accordance with the competition terms published in advance. Following the testing, we reset the competition. We began accepting submissions for the private phase on July 1, shortly after the conclusion of the public phase. Private phase submissions were evaluated as they were received until August 22, after which we announced the results and winners on August 25.

Nineteen teams submitted solutions to the Python track and eight teams submitted to the Kotlin track during the public phase. Only six teams participated in the private phase. Winners of the competition received monetary prizes and license grants from JetBrains, API keys from Mistral~AI, and certificates of achievement. 

We also invited all participating teams to present their solutions at the designated workshop session at the 40th IEEE/ACM International Conference on Automated Software Engineering (ASE 2025). Five out of the six teams submitted papers, which were subsequently reviewed in September. We evaluated and checked all the submissions, after which we added all authors to the PC and asked them to provide feedback about each others' papers. Such a light-weight cross-review process allowed the authors to suggest valuable clarifications to the papers, as well as to familiarize contestants with other strategies. In the end, quite a lot of feedback was provided, and we accepted all five papers, in addition to publishing this report.

\begin{table}[t]
\centering
\caption{\label{tab:timeline}Competition timeline (all dates are in 2025).}
\begin{tabular}{lc}\toprule
\textbf{Activity} & \textbf{Date} \\\midrule
Internal testing starts & April 14 \\
Internal testing ends & April 25 \\
Workshop acceptance notification & April 25 \\
Public phase starts & June 9 \\
Public phase ends & July 25 \\
Private phase starts & July 25 \\
Paper submission starts & July 25 \\
Private phase ends & August 22 \\
Announcement of final results & August 25 \\
Paper submission ends & September 1 \\
Papers quality checked & September 12 \\
Cross-review starts & September 12 \\
Notifications sent & September 26 \\
Camera-ready papers & October 5 \\
Workshop session & November 18 \\\bottomrule
\end{tabular}
\end{table}

\section{\label{sec:solutions}Solutions}

Six teams submitted their solutions to the private phase of the competition. After analyzing the solutions, we observed that most strategies share several common traits. They use a parsing tool to extract complete definitions and the symbols appearing in the prefix and suffix. They apply a classical information retrieval ranking method such as BM25, combined with a number of heuristics, to retrieve the highest-scoring code chunks as the resulting context, which is then trimmed to match the context size of the upstream LLM. We present our summaries of the submitted solutions in the following subsections, with references to papers where possible.

Tables~\ref{tab:python:public} and \ref{tab:kotlin:public} represent the team standings during the public Python and Kotlin phases, respectively. Tables~\ref{tab:python:private} and \ref{tab:kotlin:private} represent the final standings on the private Python and Kotlin phases, respectively.

\begin{table*}[t]
\centering
\caption{\label{tab:python:public}Team standings on the Python public phase of the competition. The main metric is Average chrF, higher is better. The leaderboard is available at \url{https://eval.ai/web/challenges/challenge-page/2516/leaderboard/6298}.}
\begin{tabular}{clcccc}
\toprule
\textbf{Rank} & \textbf{Team} & \textbf{Average chrF $\downarrow{}$} & \textbf{Mellum chrF} & \textbf{Codestral chrF} & \textbf{Qwen chrF} \\
\midrule
1 & NoMoreActimel & 0.7469 & 0.6950 & 0.8034 & 0.7424 \\
2 & SpareCodeComplete & 0.6152 & 0.5563 & 0.6568 & 0.6324 \\
3 & REALISE Lab & 0.5586 & 0.5230 & 0.6032 & 0.5496 \\
4 & WSPR\_NCSU & 0.5565 & 0.4993 & 0.6128 & 0.5574 \\
5 & init commit & 0.5510 & 0.5150 & 0.5969 & 0.5411 \\
6 & deepto98 & 0.5502 & 0.5083 & 0.6144 & 0.5277 \\
7 & Aleksei Solovev & 0.5416 & 0.5059 & 0.5819 & 0.5369 \\
8 & wanteatfruit & 0.5386 & 0.5031 & 0.5838 & 0.5287 \\
9 & SaNDwich\&TEST & 0.5372 & 0.5076 & 0.5728 & 0.5311 \\
10 & Prometheus-Agent & 0.5339 & 0.4981 & 0.5689 & 0.5348 \\
11 & Chef Emoji & 0.5308 & 0.5017 & 0.5689 & 0.5218 \\
12 & Vanilla & 0.5282 & 0.4976 & 0.5725 & 0.5147 \\
13 & StarAtNyte & 0.5238 & 0.4946 & 0.5625 & 0.5143 \\
14 & Piotr Kasprowicz (bm25) & 0.5236 & 0.4877 & 0.5581 & 0.5250 \\
15 & Biznismeni & 0.5230 & 0.4877 & 0.5562 & 0.5251 \\
16 & miss MISIS & 0.5225 & 0.4877 & 0.5546 & 0.5251 \\
17 & Someone & 0.5219 & 0.4928 & 0.5567 & 0.5161 \\
18 & SA Team & 0.5199 & 0.4902 & 0.5496 & 0.5198 \\
--- & Baseline: Recent Files & 0.5172 & 0.4868 & 0.5605 & 0.5042 \\
--- & Baseline: Random & 0.5126 & 0.4942 & 0.5447 & 0.4990 \\
19 & rhythm2211 (test) & 0.2985 & 0.1204 & 0.3936 & 0.3814 \\
\bottomrule
\end{tabular}
\end{table*}

\begin{table*}[t]
\centering
\caption{\label{tab:kotlin:public}Team standings on the Kotlin public phase of the competition. The main metric is Average chrF, higher is better. The leaderboard is available at \url{https://eval.ai/web/challenges/challenge-page/2516/leaderboard/6299}.}
\begin{tabular}{clcccc}
\toprule
\textbf{Rank} & \textbf{Team} & \textbf{Average chrF $\downarrow{}$} & \textbf{Mellum chrF} & \textbf{Codestral chrF} & \textbf{Qwen chrF} \\
\midrule
1 & SpareCodeComplete & 0.7125 & 0.6791 & 0.7442 & 0.7143 \\
2 & init commit & 0.6786 & 0.6681 & 0.7092 & 0.6586 \\
3 & WSPR\_NCSU & 0.6607 & 0.6015 & 0.7160 & 0.6645 \\
4 & NoMoreActimel & 0.6590 & 0.6115 & 0.7139 & 0.6515 \\
5 & REALISE Lab & 0.6577 & 0.6336 & 0.7048 & 0.6346 \\
6 & deepto98 & 0.6555 & 0.6147 & 0.7060 & 0.6457 \\
7 & Wu Wei & 0.6425 & 0.6237 & 0.6783 & 0.6253 \\
8 & SaNDwich\&TEST & 0.6373 & 0.6287 & 0.6772 & 0.6061 \\
--- & Baseline: Recent Files & 0.6351 & 0.6201 & 0.6702 & 0.6150 \\
--- & Baseline: Random & 0.6274 & 0.6201 & 0.6598 & 0.6024 \\
\bottomrule
\end{tabular}
\end{table*}

\subsection{Team SaNDwich\&TEST: extraction of definitions for symbols mentioned in the completion point}

The team first enumerated the imported files and added them as candidates while limiting their sizes. They then extracted all symbol names from the prefix and suffix and retrieved their definitions, using the \texttt{ast} module for Python and regular expressions for Kotlin as a fallback strategy. If the candidate list was empty, they selected a random file from the modified set. Finally, they assembled the context by concatenating the resulting snippets. At the private phase of the competition, this team was ranked 5th on Python and 4th on Kotlin.

\subsection{Team Wu Wei: heuristical ranking of code snippets from the Kotlin PSI representation}

The team parsed the entire project into a program structure interface (PSI) tree,\footnote{PSI: \url{https://plugins.jetbrains.com/docs/intellij/psi.html}} distinguishing functions, classes, objects, and variables. They then generated candidate contexts by resolving symbols from the prefix and suffix and pulling declarations from modified files. Next, they ranked the candidates using heuristics based on signals such as current file declarations, package distance, typing, and incoming references. Finally, they assembled the context using PSI formatters within the token budget. Due to the use of the tooling specific to the Kotlin programming language, this team participated only in the Kotlin track, in which they were ranked 5th at the private phase. You can find more details in the authors' paper~\cite{WuWei}.

\subsection{Team WSPR\_NCSU: retrieval of augmented code chunks using hybrid search}

The team split all project files into overlapping line-based chunks. They then built BM25 and FAISS \cite{Johnson:19} indices, using the all-MiniLM-L6-v2\footnote{Model: \url{https://huggingface.co/sentence-transformers/all-MiniLM-L6-v2}} encoder model for embeddings~\cite{Reimers:19}. Next, they retrieved the top-$k$ chunks for both prefix and suffix, fused the BM25 and FAISS results, and augmented them with adjacent context from the original files. The augmented contexts were obtained by fetching the chunks of code in the file with the completion point that immediately follows the prefix in the original file (and vice versa for the suffixes). Finally, they assembled the context from three sources: the full content of the file containing the completion point, the modified files, and the augmented retrieved chunks. Our repository snapshots did not disclose the ground truth, so this approach for assembling does not leak any ground truth to the context. At the private phase of the competition, this team was ranked 4th on Python and 3rd on Kotlin, sharing the rank with the team REALISE Lab on Kotlin due to the negligible difference in the final scores. You can find more details in the authors' paper~\cite{WSPR_NCSU}.

\begin{table*}[t]
\centering
\caption{\label{tab:python:private}Team standings on the Python private phase of the competition. The main metric is Average chrF, higher is better. Teams NoMoreActimel, SpareCodeComplete, and REALISE Lab received the prizes.}
\begin{tabular}{clcccc}\toprule
\textbf{Rank} & \textbf{Team} & \textbf{Average chrF $\downarrow{}$} & \textbf{Mellum chrF} & \textbf{Codestral chrF} & \textbf{Qwen chrF} \\\midrule
 1 & \textbf{NoMoreActimel} & \textbf{0.734} & 0.656 & 0.820 & 0.725 \\
 2 & \textbf{SpareCodeComplete} & \textbf{0.725} & 0.695 & 0.766 & 0.713 \\
 3 & \textbf{REALISE Lab} & \textbf{0.644} & 0.613 & 0.710 & 0.608 \\
 4 & WSPR\_NCSU & 0.636 & 0.582 & 0.710 & 0.615 \\
 --- & Baseline: BM25 & 0.610 & 0.585 & 0.659 & 0.585 \\
 5 & SaNDwich\&TEST & 0.610 & 0.590 & 0.661 & 0.578 \\
 --- & Baseline: Recent & 0.606 & 0.576 & 0.657 & 0.587 \\\bottomrule
\end{tabular}
\end{table*}

\begin{table*}[t]
\centering
\caption{\label{tab:kotlin:private}Team standings on the Kotlin private phase of the competition. The main metric is Average chrF, higher is better. Teams SpareCodeComplete, NoMoreActimel, WSPR\_NCSU, and REALISE Lab received the prizes.}
\begin{tabular}{clcccc}\toprule
\textbf{Rank} & \textbf{Team} & \textbf{Average chrF $\downarrow{}$} & \textbf{Mellum chrF} & \textbf{Codestral chrF} & \textbf{Qwen chrF} \\\midrule
 1 & \textbf{SpareCodeComplete} & \textbf{0.748} & 0.723 & 0.769 & 0.753 \\
 2 & \textbf{NoMoreActimel} & \textbf{0.731} & 0.684 & 0.791 & 0.719 \\
 3 & \textbf{WSPR\_NCSU} & \textbf{0.660} & 0.616 & 0.709 & 0.653 \\
 3 & \textbf{REALISE Lab} & \textbf{0.659} & 0.652 & 0.688 & 0.637 \\
 4 & SaNDwich\&TEST & 0.635 & 0.633 & 0.658 & 0.613 \\
 --- & Baseline: BM25 & 0.634 & 0.627 & 0.652 & 0.621 \\
 5 & Wu Wei & 0.627 & 0.624 & 0.648 & 0.609 \\
 --- & Baseline: Recent & 0.620 & 0.618 & 0.636 & 0.605 \\\bottomrule
\end{tabular}
\end{table*}

\subsection{Team REALISE Lab: retrieval of Tree-sitter-based code chunks using BM25}

The team first transformed the prefix and suffix into trimmed queries by extracting symbol definitions with Tree-sitter.\footnote{Tree-sitter: \url{https://tree-sitter.github.io/tree-sitter/}} They then identified the last top-level block before the completion point and the next block starting afterward. Next, they created a synthetic corpus of augmented function definitions, where each entry included the full function context with parent classes and annotations. Finally, they retrieved the top-$k$ entries using BM25. At the private phase of the competition, this team was ranked 3rd on Python and 3rd on Kotlin, sharing the rank with the team WSPR\_NCSU on Kotlin due to the negligible difference in the final scores. You can find more details in the authors' paper~\cite{REALISELab}.

\subsection{Team NoMoreActimel: query reformulation for retrieval-augmented generation with item boosting}

The team chunked the code and embedded the chunks with the Qwen3-Embedding-0.6B\footnote{Model: \url{https://huggingface.co/Qwen/Qwen3-Embedding-0.6B}} model to build a FAISS index, using the \texttt{ast} module for Python and overlapping character-based sliding windows for Kotlin. They then created queries from completion points using multiple strategies: full file as a query, queries in the form of fixed-size chunks, query of $N$ lines around the completion point, and their versions with textual explanations of code generated by Qwen2.5-Coder-1.5B-Instruct\footnote{Model: \url{https://huggingface.co/Qwen/Qwen2.5-Coder-1.5B-Instruct}} appended as suffixes. Next, the authors ranked queries according to multiple heuristics based on query location in the file and query length. Then, they embedded the queries and performed retrieval-augmented generation using cosine similarity with heuristic boosting, excluding low-scoring candidates. Finally, they assembled the context iteratively, counting tokens with the Mellum-4b-sft-python tokenizer.\footnote{Model: \url{https://huggingface.co/JetBrains/Mellum-4b-sft-python}} At the private phase of the competition, this team was ranked 1st on Python and 2nd on Kotlin. You can find more details in the authors' paper~\cite{NoMoreActimel}.

\subsection{Team SpareCodeComplete: querying the trigram index using symbols extracted from Tree-sitter}

The team used manually-written Tree-sitter grammars to extract functions, classes, variables, and other important symbols in code. They generated queries to the Zoekt\footnote{Zoekt: \url{https://github.com/sourcegraph/zoekt}} code search engine by combining AST nodes with the parsed symbols. For constructing queries, they used information about the difference between the current file state and the previous revision. Next, they ranked the extracted symbols by their distance to the completion point and substituted the top-ranked ones into disjunctive query clauses. Zoekt then handled both retrieval and ranking using signals such as content, filenames, substrings, and boundaries. Finally, they assembled the context by concatenating the top-$k$ results from Zoekt. At the private phase of the competition, this team was ranked 2nd on Python and 1st on Kotlin. You can find more details in the authors' paper~\cite{SpareCodeSearch}.

\subsection{Threats to Validity}

For building the search indices, the solutions by teams SpareCodeComplete and NoMoreActimel use all available data in the provided dataset, including other repositories and all the revisions for the given repository. Such an approach may lead to a data leakage from the future versions of the same repository, as they may contain the reference code snippet. It poses a threat to the validity of the results, which has to be investigated further.

\section{\label{sec:conclusion}Conclusion}

The competition we organized as part of the ASE~2025 conference allowed us to explore the solution space and identify the most practical strategies for context collection. The competition was conducted in three phases: practice, public, and private, with the final phase used to determine the official rankings. During the public phase, we attracted nineteen teams to the Python track and eight teams to the Kotlin track, with only six teams in total submitting code for the private phase. Participants had no influence over the completions, only over the context they collected, which affected the performance of three strong code completion models: Codestral, Qwen2.5-Coder, and Mellum. The solutions employed practical combinations of parsing and retrieval to assemble highly relevant contexts. We released the complete competition dataset on Zenodo~\cite{Dataset}. We believe that the insights and experience gained from this competition will provide the research community with opportunities to improve the coding experience for millions of developers worldwide.

\section*{Acknowledgments}

We would like to express our gratitude to our partners at Mistral~AI: Baptiste Rozi\`{e}re and Sophia Yang, as well as to our colleagues at JetBrains: Nikolai Baranko, Timofey Bryksin, Kirill Chekmenev, Ozren Dabi\'{c}, Valentin Fondaratov, Ekaterina Frolova, Valerie Kuzmina, Artem Mukhin, Semyon Proshev, Patrycja Sztorc, and Vladislav Tankov. We are also grateful to the workshop chairs of ASE~2025 for providing the opportunity to host the competition as part of the conference. Finally, we would like to thank all the participants in the competition.

\bibliographystyle{IEEEtran}
\balance
\bibliography{report}

\end{document}